\documentclass{emulateapj}
\usepackage{epsfig}
\usepackage{natbib}

\newcommand{\h}{$^h$}
\newcommand{\m}{$^m$}
\newcommand{\s}{$^s$}
\newcommand{\one}{$J=1\rightarrow0$}
\newcommand{\two}{$J=2\rightarrow1$}
\newcommand{\three}{$J=3\rightarrow2$}
\newcommand{\four}{$J=4\rightarrow3$}

\newcommand{\six}{$J=6\rightarrow5$}
\newcommand{\seven}{$J=7\rightarrow6$}
\newcommand{\nine}{$J=9\rightarrow8$}
\newcommand{\da}{$\Delta\alpha$}
\newcommand{\dd}{$\Delta\delta$}
\newcommand{\kms}{km~s$^{-1}$}

\newcommand{\cc}{cm$^{-3}$}
\newcommand{\dndv}{cm$^{-2}$~s~km$^{-1}$}
\newcommand{\Ta}{T$^*_A$}
\newcommand{\ki}{$\chi^2$}
\newcommand{\kir}{$\chi^2_r$}
\newcommand{\coTW}{$^{12}$CO}
\newcommand{\coTH}{$^{13}$CO}
\defcitealias{wil01}{WMKH01}

\begin{document}

\title{A Map of OMC-1 in CO J$=9\rightarrow8$}
\author{Daniel~P.~Marrone\altaffilmark{1},  
James~Battat\altaffilmark{1}, Frank~Bensch\altaffilmark{1,2},
Raymond~Blundell\altaffilmark{1}, Marcos~Diaz\altaffilmark{3},
Hugh~Gibson\altaffilmark{1,4}, Todd~Hunter\altaffilmark{1},
Denis~Meledin\altaffilmark{1,5}, Scott~Paine\altaffilmark{1},
D.~Cosmo~Papa\altaffilmark{1}, Simon~Radford\altaffilmark{6},
Michael~Smith\altaffilmark{1}, and Edward~Tong\altaffilmark{1}}
\altaffiltext{1}{Harvard-Smithsonian Center for Astrophysics, 
	60 Garden Street, Cambridge, MA 02138;
	dmarrone@cfa.harvard.edu (corresponding author),
	jbattat@cfa.harvard.edu, rblundell@cfa.harvard.edu,
	thunter@cfa.harvard.edu, spaine@cfa.harvard.edu,
	cpapa@cfa.harvard.edu, msmith@cfa.harvard.edu,
	etong@cfa.harvard.edu.}
\altaffiltext{2}{Current address: Radioastronomisches Institut der
Universit\"{a}t Bonn, Auf dem H\"{u}gel 71, D-53121 Bonn, Germany;
fbensch@astro.uni-bonn.de.}
\altaffiltext{3}{Boston University, Department of Electrical 
	and Computer Engineering, 8 St. Mary's Street, Boston, MA
	02215; mardiaz@bu.edu.}
\altaffiltext{4}{Current address: RPG Radiometer Physics GmbH,
	Birkenmaarstrasse 10, 53340 Meckenheim, Germany;
	hgibson@cfa.harvard.edu.}
\altaffiltext{5}{Current address: GARD/MC2, Fysikgrand 3,
	Chalmers University of Technology, S-412 96 Gothenburg,
	Sweden; meledin@oso.chalmers.se.}
\altaffiltext{6}{National Radio Astronomy Observatory, 
	949 North Cherry Avenue, Tucson, AZ 85721; sradford@nrao.edu}

\begin{abstract}
The distribution of $^{12}$C$^{16}$O {\nine} (1.037~THz) emission has
been mapped in OMC-1 at 35 points with 84{\arcsec} resolution. This is
the first map of this source in this transition and only the second
velocity-resolved ground-based observation of a line in the THz
frequency band. There is emission present at all points in the map, a
region roughly 4{\arcmin} $\times$ 6{\arcmin} in size, with peak
antenna temperature dropping only near the edges. Away from the
Orion~KL outflow, the velocity structure suggests that most of the
emission comes from the OMC-1 photon-dominated region, with a typical
line width of 3-6~{\kms}. Large velocity gradient modeling of the
emission in {\nine} and six lower transitions suggests that the lines
originate in regions with temperatures around 120~K and densities of
at least $10^{3.5}$~{\cc} near $\theta^{1}$C Ori and at the Orion bar,
and from 70~K gas at around $10^4$~{\cc} southeast and west of the
bar. These observations are among the first made with the 0.8~m
Smithsonian Astrophysical Observatory Receiver Lab Telescope, a new
instrument designed to observe at frequencies above 1~THz from an
extremely high and dry site in northern Chile.
\end{abstract}


\section{INTRODUCTION}
The \objectname{OMC-1} region of the \objectname{Orion A} molecular
cloud is the nearest site ($\sim$500~pc) of recent high mass star
formation. Its diverse components make it a natural laboratory for the
study of many stages of star formation. The OMC-1 cloud is one of
three clouds connected by the Orion ``ridge,'' a long and dense
molecular filament \citep{cas90}. The \ion{H}{2} regions M42 and M43
sit in front of the ridge, and extended narrow-line CO emission (the
``spike'' component) is excited in the photon-dominated region (PDR)
at the interface between the ionized and neutral gas. The brightest
star in the Trapezium, \objectname{$\theta^{1}$C Ori}, is the dominant
source for the far-ultraviolet photons that sustain the M42 \ion{H}{2}
region and its associated PDR \citep{sta93}. One arcminute to the
north-west of $\theta^{1}$C~Ori, the Kleinmann-Low nebula (KL) is the
brightest source outside the solar system at 20~$\mu$m, with a
luminosity of $\sim10^5 L_\odot$ \citep{KL}. Several powerful sources
have been identified in the infrared within KL, including the
Becklin-Neugebauer point source \citep[BN;][]{BN}, likely a runaway B
star \citep{plam95,tan04}, and IRc2, once thought to be the power
source for the nebula \citep{down81} but more recently resolved into
several sources that may not be self-luminous
\citep{doug93,MR95,geza98}. Molecular line observations of
\objectname{Orion KL} show emission extending out to nearly
$\pm$100~{\kms} \citep[e.g., CO {\seven} and {\four}
in][hereafter WMKH01]{wil01}. This large line width, usually called
``plateau'' emission, is due to an outflow long attributed to IRc2
\citep{wrig83} that is more likely associated with nearby radio and
infrared point sources \citep{MR95,gree98}. The Orion bar is a well
studied region of the PDR south-east of the Trapezium. At this
location the ionization front is viewed edge-on over a length of
$\sim$3{\arcmin}, allowing direct examination of the chemical
stratification in a plane-parallel PDR \citep[e.g.,][]{werf96,walm00}.

The OMC-1 cloud has been mapped in many millimeter, submillimeter, and
far-IR transitions (e.g., \citealt{sta93,gol97,her97,plu00,sem00,iked02};
\citetalias{wil01}). Previous observations in CO transitions up to
{\seven} (\citealt{sch95}; \citetalias{wil01}) have found warm PDR gas
extending over a very large region. In this paper we present the first
map of OMC-1 in the {\nine} transition of $^{12}$C$^{16}$O. This line
has previously been observed in OMC-1 from the Kuiper Airborne
Observatory \citep{roe91,betz93} and from the Heinrich Hertz
Submillimeter Telescope \citep{kaw02}, but never at more than a few
points. Our map contains 35 pointings and covers $\sim$25 arcmin$^2$
at 84{\arcsec} resolution. Our spectra provide information about the
population of the $J=9$ rotational level (250~K), nearly 100~K higher
than the $J=7$ level probed by {\seven} maps, and the critical density
at {\nine} ($\sim10^6$~\cc) is approximately twice that at {\seven}
($\sim5\times10^5$~\cc). This THz transition should therefore be a
better probe of the density and temperature of warm and dense sources
such as OMC-1, where lower transitions are thermalized.

Because of absorption by the Earth's atmosphere, the {\nine} transition of
CO as well as other astronomically interesting lines at frequencies
above 1~THz are unobservable from current ground-based observatories
except in very unusual weather. However, recent atmospheric
measurements have demonstrated that several windows can open up
between 1 and 3~THz at very high and dry locations
\citep{pai00,mat99}. Under favorable conditions, three of these
windows, centered at 1.03, 1.35, and 1.5~THz, show transmission as high
as 40\%. These windows include six rotational transitions of {\coTW}
and {\coTH}, as well as the ground-state fine-structure transition of
singly ionized nitrogen, one of the most important coolants of the
diffuse ISM, and numerous other common and exotic atomic and molecular
lines. A new telescope, the Receiver Lab Telescope (RLT), has been
constructed by the Smithsonian Astrophysical Observatory to take
advantage of these windows, and is currently deployed at 5525~m
altitude in northern Chile. This map of OMC-1 represents some of the
first observations made with the RLT and demonstrates the capabilities
of this new instrument.

\section{OBSERVATIONS}

We observed the OMC-1 region of the Orion A molecular cloud in the
{\nine} transition of $^{12}$C$^{16}$O using the 0.8~m diameter RLT
\citep{blu02} on Cerro Sairecabur, Chile. At the 1036.912~GHz
frequency of this line we expect a beamsize of 84{\arcsec}. Our 330
channel autocorrelation spectrometer has a bandwidth of 286~{\kms}
(990~MHz) and a resolution of 1.04~{\kms} at this frequency. The data
were acquired on the morning of 2002 December 15 between 05:30 and
07:30~UT. We obtained spectra at 35 points with 50{\arcsec} spacing in
an irregularly shaped region of OMC-1. These observations were made in
position-switching mode with a reference point 10{\arcmin} to the west
of each map point. Each spectrum is the sum of 12 integration cycles
of 5~s on-source and 5~s off-source. The map is centered on
$\alpha$~=~05\h35\m14{\s} and
$\delta$~=~$-$05\degr22\arcmin31{\arcsec} (J2000.0), which coincides
with the location of several Orion line surveys
\citep{sut85,bla86,sch01}. Orion KL, BN, and IRc2 are all well within
the beam at this location. The ranges in offsets from the map center
are \da~=~$-$50{\arcsec} to +150{\arcsec} and \dd~=~$-$250{\arcsec} to
+100\arcsec.

The telescope pointing varies from night to night by up to tens of
arcseconds because of shifts in the telescope foundation. To remove
these variations we used an optical guidescope to determine offsets
from our base pointing model. The limitation of this method is the
accuracy with which we know the offset between the optical and radio
beams, about 15{\arcsec}, and we take this value to be our absolute
pointing error. Within the map the relative pointing is considerably
better. To ensure that there was no systematic drift in the (0\arcsec,
0\arcsec) point over the observation period we remeasured the pointing
offsets on a nearby star approximately every 25 minutes. We also
measured the center spectrum twice; the two measurements were
separated by 1~hr, and there were three measurements of the pointing
offsets made during that interval. As a check of the consistency of
these two spectra, we rescale them so that the averages of their three
highest channels are the same and compute the rms difference between
them in the nonbaseline channels. When the spectra are scaled to 100~K
we measure an rms difference of 4.8~K, very similar to the 4.2~K we
expect from the noise seen in the baseline channels, and the
difference spectrum shows no remnant structure.  Adjacent spectra are
significantly different; the most similar spectrum, at (0\arcsec,
$-$50\arcsec), differs from the two center-point spectra by 7.7 and
7.4~K and has narrower wings. We also note that the averages of the
three highest channels are 65.5 and 64.1~K for the first and second
center-point spectra, respectively, indicating that the calibration
was self-consistent to 2\% over this interval.

The hot electron bolometer receiver used for these measurements is
described elsewhere \citep{mel04}. During these observations the
double-sideband receiver temperature was found to be 980~K through
observations of cold (liquid nitrogen) and ambient loads. Atmospheric
transmission measurements were made every 10 minutes by a
Fourier transform spectrometer \citep{pai00} located a few meters away
from the telescope. Over the course of the observations the zenith
transmission at the line frequency decreased from 23.5\% to 18\%. To
determine a single-sideband system temperature (T$_{sys}$), we assume
equal gain in the signal and image sidebands (centered on 1036.912 and
1030.848~GHz, respectively) because the mixer frequency response is
essentially flat between 900~GHz and 1.05~THz. Our data are calibrated
to the {\Ta} scale of \cite{uli76}. We have measured a beam efficiency
\citep[$\eta_l$ in][]{uli76} of 45\% at 883~GHz by mapping a telluric
ozone absorption feature against Jupiter (angular diameter 42\arcsec;
\citealt{marr04}). This efficiency is much lower than expected; the
RLT primary mirror has an rms figure error of only 3~$\mu$m, causing a
2\% Ruze loss, and our blockage and illumination pattern give a
theoretical efficiency of 94\%. We believe that this indicates that
the receiver was poorly aligned with the telescope and that a
significant portion of the receiver beam was being truncated before
reaching the secondary mirror. We have not made this beam measurement
nearer to the CO {\nine} frequency, so we adopt this efficiency for
the calibration. After accounting for the above factors, the
single-sideband system temperature at the source elevation increased
from $2.7\times10^4$ to $6.4\times10^4$~K during our observations as
the transmission degraded and Orion set. From estimates of the
uncertainties in the atmospheric transmission and the efficiency, we
expect that our calibration is accurate to $\pm$25\%.

\section{RESULTS}

\begin{figure}
\epsscale{1.1}
\plotone{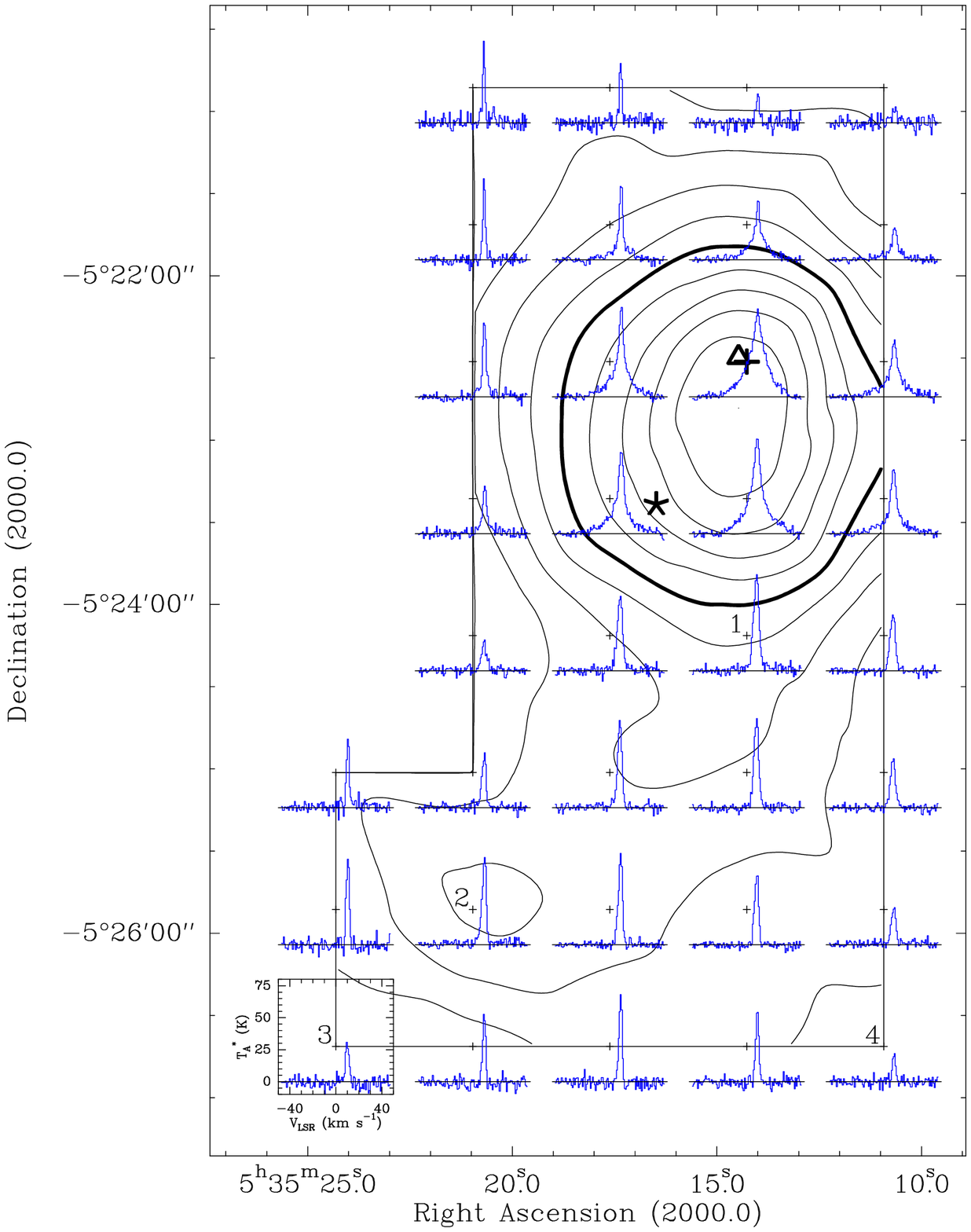}
\caption{CO {\nine} emission in OMC-1. The contours trace the line
flux between $-$30 and +40~{\kms} (in K~\kms). The contours are spaced
by 10\% of the peak flux (1260~K~{\kms} near the center point) and run
from 10 to 90\%, with the 50\% contour reinforced. The small crosses
mark the location of map pointings, and the bold cross marks the
center point of the map at $\alpha$~=~05\h35\m14{\s},
$\delta$~=~$-$05\degr22\arcmin31{\arcsec} (J2000.0). The individual
spectra lie on a square grid with 50{\arcsec} spacing. The star marks
the position of $\theta^{1}$C~Ori, and the triangle marks the position
of the IRc2 complex of sources. The numbers 1-4 correspond to the
positions in Tables~\ref{data} and \ref{fits}. The spectrum at each
map point is also plotted. The scale for each spectrum is the same as
that given in the lower left.}
\label{map}
\end{figure}

The calibrated spectrum at each of the 35 map points is shown in
Figure~\ref{map} superposed on a contour map of the CO {\nine} line
flux integrated between $-$30 and +40~\kms. The narrow spike emission
from the PDR is evident throughout the map, while broad plateau
emission is only observed near the center, around Orion KL. Linear
baselines were fitted to 27 channels ($-$47 to $-$30~{\kms} and +40 to
+47~\kms) in each spectrum and removed. Because of mismatched
baselines in the output of the three independent sections of our
autocorrelation spectrometer, it was only possible to use the middle
110 channels (covering $-$47 to +47~\kms) in our final
spectra. Unfortunately, other observations of Orion~KL in this
transition \citep{roe91,betz93,kaw02} and lower transitions (e.g.,
\citealt{sch95}; \citetalias{wil01}) show plateau emission at
velocities outside the $-$30 to +40~{\kms} range. Hence, the baselines
subtracted from the (\da~=~0\arcsec, ~=~0\arcsec) and (0\arcsec,
$-$50\arcsec) spectra, where the emission is widest in velocity, are
likely to be inaccurate, and the integrated emission obtained from
these spectra is significantly diminished.

Figure~\ref{map} shows that the peak emission in the {\nine} line is
$\sim$20{\arcsec} south of our map center. Observations of lower CO
transitions, from {\one} and {\two} \citep{cas90} to {\seven}
(\citealt{how93}; \citetalias{wil01}), show peak emission at the
location of IRc2, marked with the triangle in Figure~\ref{map}. The
addition of reasonable flux corrections at the two central points
affected by the baseline problem can move the center point toward
(0\arcsec, 0\arcsec) by as much as 10{\arcsec}. The remaining error is
consistent with our absolute pointing uncertainty of 15\arcsec.

The outflow near IRc2 is unresolved in our map, as expected from its
$\sim$50{\arcsec} size in {\seven} \citepalias{wil01}. We have used
the emission in the line wings, which is spatially confined to this
outflow, as a source to map the beam and estimate the beam size. For
this estimate we assume Gaussian profiles for the source and beam, the
simplest possible assumption.  We obtain a range of half-power widths
between 85{\arcsec} and 105{\arcsec}, which, after deconvolution with
a 50{\arcsec} source, give beam sizes between 69{\arcsec} and
92{\arcsec}. These values provide confirmation that the telescope beam
is similar to the 84{\arcsec} we predict from the telescope and
receiver optics design.

\section{DISCUSSION}

\subsection{Radiative Transfer Modeling}
CO {\nine} emission is present over the entire mapped region, and the
peak line temperature decreases only at the edges. \citetalias{wil01}
also found {\seven} emission over their entire (slightly smaller)
mapping region. As mentioned previously, we expect the {\nine} line to
be a better probe of the warm and dense OMC-1 PDR. To examine the
conditions in the PDR, we compare the emission in this line to that in
lower transitions: {\coTW} {\seven} and {\four} \citepalias{wil01},
{\six} (K.~N.~Allers 2005, in preparation), {\three}
\citep{tigg93}, {\coTH} {\three} \citep{tigg93}, and {\one}
\citep{plu00}. The data sets have been convolved to the 84{\arcsec}
resolution of the RLT, except for the {\six} data, which have 86{\arcsec}
resolution. We estimate the density and temperature at various map
points using the large velocity gradient (LVG) model of
\citet{stut85}. LVG models assume that ordered line-of-sight motions
are large compared to local random velocities and that there is a
one-to-one correspondence between position along the line of sight and
velocity, allowing the radiative transfer to be treated locally
\citep{scov74}. The free parameters in these models are: the molecular
hydrogen density [$n$(H$_2$), or simply $n$], the CO column density
per velocity interval ($dN/dv$), and the kinetic temperature
(T$_{kin}$). The model performs calculations through $J=20$ with the
H$_2$-CO collision coefficients of \citet{flow85}. Four representative
points were selected from the map (marked in Figure~\ref{map}); the
Rayleigh-Jeans main-beam brightness temperatures in each transition at
each position are given in Table~\ref{data}. To correct the {\nine}
data from {\Ta} to T$_{MB}$, we use a source coupling efficiency of
85\% because of the extended nature of the spike emission. This is the
same coupling efficiency used with the {\six} data by K.~N.~Allers
(2005, in preparation) for a similar beam size. Coupling efficiencies
for the other data are obtained from the original papers. Position 1
is near $\theta^{1}$C Ori, position 2 lies on the Bar, and positions 3
and 4 fall on the edges of the mapped region where the {\nine}
emission is decreasing.

We estimate the model parameters using a {\ki} minimization
analysis. The peak brightness temperature in each line is determined
from a Gaussian fit. These temperatures are compared to a grid of LVG
models covering T$_{kin}$ from 40 to 200~K, $n$ from $10^{2.5}$ to
$10^{6.5}$~\cc, and $dN/dv$ from $10^{15.5}$ to $10^{20}$~\dndv. We
assign an error of 25\% to each temperature to account for the
uncertainty in the inter-telescope calibration and the error-beam
contribution to the observed temperatures on this extended source. For
calculations of {\coTH} line temperatures we use a column density
ratio of $N(^{12}$CO$)/N(^{13}$CO$)=67$, as determined in Orion~A by
\citet{lang90}, and assume that the {\coTW} and {\coTH} emission arise
in gas of the same temperature. The T$_{kin}$ values we derive are
very similar to those measured for {\coTH} by \citet{plu00}, where the
gas kinetic temperature was determined from measurements of {\coTH}
lines only, suggesting that this assumption is a good approximation
for this source.

The parameters derived from these fits are given in Table~\ref{fits}.
The goodness of fit at each position is judged by examining the
minimum value of the reduced {\ki} (\kir), or {\ki} divided by the
number of degrees of freedom. The ranges in the table describe the
region in parameter space where {\kir} is less than 1 plus the minimum
value. The observed minima in {\kir} indicate good fits at all four
positions.

The fits at positions 1 and 2 provide examples of the difficulties of
determining $n$ from sets of thermalized transitions. When the {\coTW}
line temperatures at these positions are converted from the
Rayleigh-Jeans scale to the Planck scale (under the assumption that
they are optically thick), they are roughly constant, indicating LTE
emission. In this equilibrium case, the line temperatures are no
longer sensitive to density and the models can only set a lower limit
to the density of the gas, that being the density required to roughly
thermalize all of the observed transitions. It is only because the
temperatures of the two {\coTH} transitions are simultaneously matched
only in a small region of densities that these positions have density
upper bounds. Higher densities at the same column density per velocity
interval are not strongly excluded because they give only slightly
larger values of {\kir}. Positions 3 and 4 indicate the value of
high-$J$ data in dense environments. At these positions the {\nine}
emission is not thermalized, allowing strong constraints on the
density. When we apply our fitting technique to these positions
without the {\nine} information we do not obtain density upper bounds.

These results match well with previous measurements, although we are
able to place tighter constraints on the gas conditions. Based on
{\coTW} {\seven} and {\coTH} {\two}, \citet{how93} found that the CO
emission could be produced by column densities per velocity interval
in the range $10^{17.3}-10^{18.6}$~\dndv, densities of
$10^{3}-10^{5}$~\cc, and temperatures of $40-200$~K. \citet{plu00}
also used LVG models and an assumed density of $10^4$~{\cc} to find a
nearly constant {\coTH} column density per velocity interval of
$10^{16.4\pm0.3}$~{\dndv}, or $10^{18.1\pm0.3}$~{\dndv} for {\coTW}
with their assumed abundance ratio, and temperatures of
$\sim$100~K. At all four positions we find a CO column within the
range $10^{17.5}-10^{18.3}$~{\dndv} and molecular hydrogen density
$10^{3.5}-10^{4.6}$~{\cc}, although higher densities are allowed at
lower likelihood at positions 1 and 2. There are significant
temperature variations over the OMC-1 cloud, with the lowest
temperatures occurring at the two positions farthest from the strong
UV source $\theta^{1}$C Ori.

\subsection{Line Features}
\begin{figure}
\epsscale{1.1}
\plotone{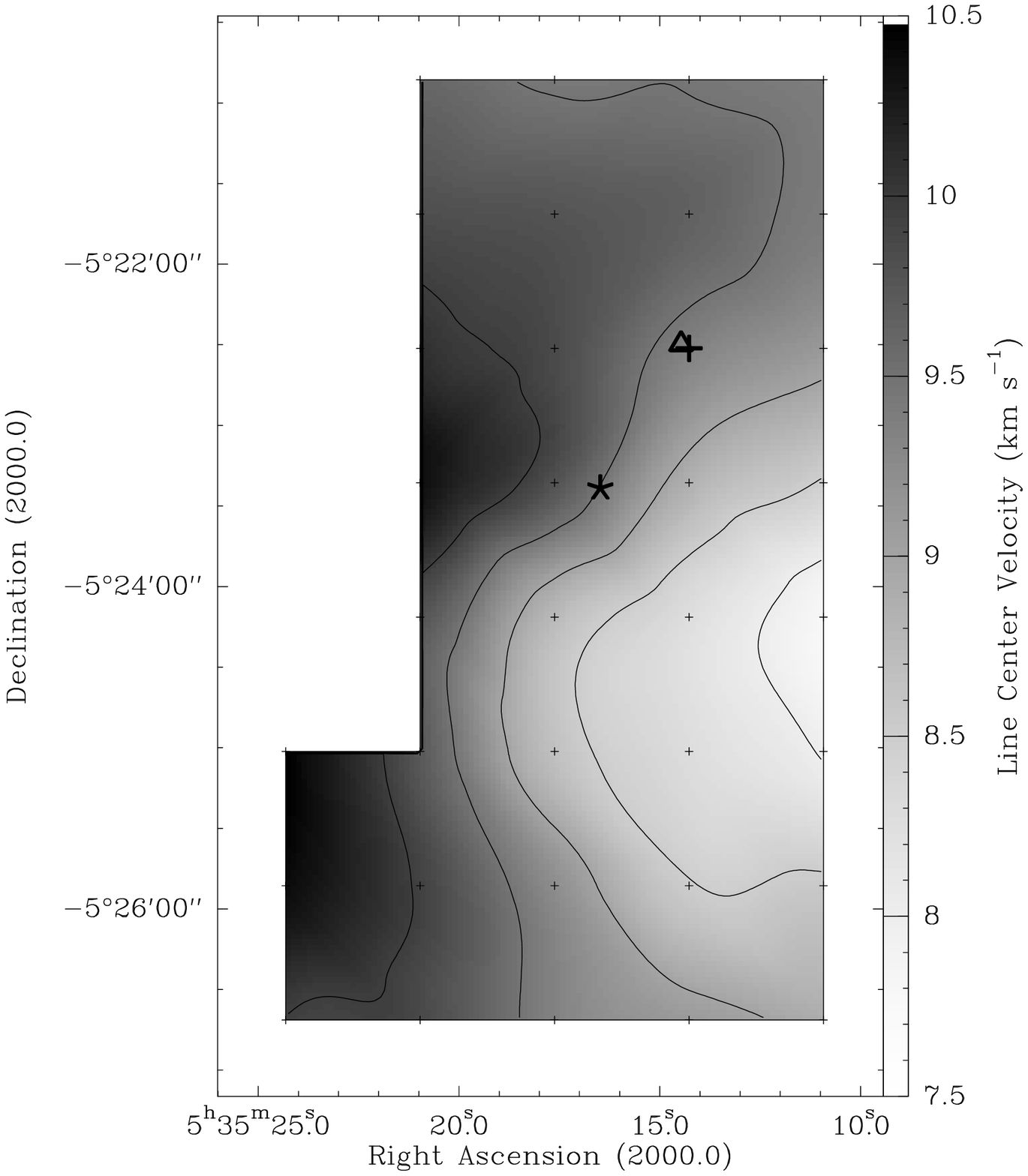}
\caption{Velocity of CO {\nine} emission in OMC-1. The line-center
velocities are obtained from each spectrum using a Gaussian fit. A
single component is used away from the KL outflow, while at the nine
center positions a broad component is added but constrained to have
the same velocity. Contours trace velocities from 8 to 10~{\kms} in
steps of 0.5~\kms. Symbols are the same as in Figure~\ref{map}.}
\label{vel_map}
\end{figure}

Away from the broad emission component around Orion~KL, the quiescent
(spike) emission dominates. Figure~\ref{vel_map} is a map of the
line-center velocities obtained from Gaussian fits to the {\nine}
spectra. At the nine positions closest to the map center where the KL
outflow is visible, we add a broad component to the fit but constrain
its velocity to match that of the narrow component. As seen in lower
transitions, such as the {\seven}, \four, and {\two} transitions in
\citetalias{wil01}, there is a gradient in the line-center velocity of
the spike emission. The minimum (least redshifted) velocity is
approximately 7.5~{\kms} near ($-$50\arcsec, $-$100\arcsec),
increasing toward the north and southeast to more than 10~{\kms} at
the eastern edge of the map. The observed velocities are
systematically redshifted (typically 0.5$-$1~\kms) from velocities
measured in the dense ridge gas \citep[e.g., CN emission
in][]{rodr01}. We interpret this as evidence that most of the CO
{\nine} emission arises in the PDR rather than gas deeper in the
molecular cloud. The redshift between the PDR surface and the ridge
may be due to the slow expansion of the \ion{H}{2} region into the
molecular cloud. Emission from the ridge, if it is present, could be
visible as a blue shoulder in individual spectra, but the only
spectrum that may show this second component is (0\arcsec,
$-$200\arcsec).

The FWHM of the spike emission, as measured from the Gaussian fits,
narrows away from the map center, decreasing from 6~{\kms} (FWHM) near
the KL outflow to about 3~{\kms} in the southeast corner and
2.5~{\kms} in the northeast. \citetalias{wil01} shows the line width
increasing to the west at the declination of {$\theta^{1}$C Ori in
{\seven}, but our angular resolution does not allow us to disentangle
the KL outflow and compare directly.

Only one spectrum shows a clear emission component other than the
spike and plateau at our resolution and signal-to-noise ratio. At the
Orion bar [(100\arcsec, $-$200\arcsec); position 2 above], a blue
component stands out above 3~{\kms}. The same component may also be
present at (100\arcsec, $-$150\arcsec) and (50\arcsec,
$-$200\arcsec). The emission at this location appears to be the sum of
two components: one around 10.5~{\kms} (consistent with the velocity
at the Bar), and another around 7.5~{\kms}; both components are
visible in the other {\coTW} and {\coTH} transitions as well. The
{\coTW} and {\coTH} {\one} channel maps of \citet{taub94} resolve the
two components in space and velocity, with the 7.5~{\kms} component
centered approximately 30{\arcsec} north and west of the bar
emission. When convolved to the same angular resolution (84\arcsec),
the line shape is very similar in the {\four} through {\nine}
transitions, suggesting that the line-center and wing components are
at similar average density and temperature and differ only in beam
filling factor. We suggest that the blue component may represent
molecular gas left over from the disrupted near edge of the molecular
cloud.

\section{CONCLUSIONS}

We have mapped CO {\nine} emission in the OMC-1 region of the Orion~A
molecular cloud over a 25 arcmin$^2$ region at 84{\arcsec}
resolution. The line is detected over the entire map, similar to what
has been seen in other submillimeter CO transitions. The high critical
density of this transition and the high energy of the $J=9$ level make
it more difficult to excite than lower levels, allowing us to measure
the molecular hydrogen density, temperature, and CO column density per
velocity interval of the gas at four locations in the PDR by combining
these spectra with lower transitions. We find that near the center of
the map even this transition may not be high enough to strongly
constrain the density because of the high gas temperatures. A wider
survey of the cloud in this line or the higher rotational lines of
{\coTW} and {\coTH} available to the RLT should reveal more cool gas
for which the THz CO transitions can tightly constrain the conditions.

This map demonstrates that it is possible to make reliable
observations from the ground in a region of the electromagnetic
spectrum normally considered to be inaccessible. In the near future we
hope to begin observations at higher frequencies to make use of the
other atmospheric windows available from our site.

\acknowledgements
We would like to thank Leonardo~Bronfman and Jorge~May of the
Universidad de Chile for invaluable support and assistance with this
project. We also thank Joseph~Salah and Mike~Poirier of the MIT
Haystack Observatory for support during the assembly and testing of
the RLT at the Westford Radio Telescope. We are grateful for the
continuous support we have received from the Director of SAO,
Irwin~Shapiro. K.~N.~Allers, A.~L.~Betz, and T.~W.~Wilson have been
kind enough to provide us with access to their data. We would like to
thank the anonymous referee for many improvements to the paper. D. P. M.
acknowledges support from an NSF Graduate Research Fellowship.

\begin{deluxetable}{cccccccc}
\tablecolumns{8}
\tablewidth{0pt}
\tablecaption{Line peak temperatures used in LVG model fitting\label{data}}
\tablehead{\colhead{Position\tablenotemark{a}} &
\colhead{T$^{12}_{9\rightarrow8}$\tablenotemark{b}} &
\colhead{T$^{12}_{7\rightarrow6}$\tablenotemark{c}} & 
\colhead{T$^{12}_{6\rightarrow5}$\tablenotemark{d}} &
\colhead{T$^{12}_{4\rightarrow3}$\tablenotemark{c}} & 
\colhead{T$^{12}_{3\rightarrow2}$\tablenotemark{e}} &
\colhead{T$^{13}_{3\rightarrow2}$\tablenotemark{e}} &
\colhead{T$^{13}_{1\rightarrow0}$\tablenotemark{f}}}
\startdata
1 &  85$\pm$21  &  137$\pm$34  &  83$\pm$21  & 
	102$\pm$26  &  93$\pm$23  &  24$\pm$6  &  11$\pm$3 \\
2 &  80$\pm$20  &  120$\pm$30  &  \nodata    &
	102$\pm$26  &  86$\pm$22  &  24$\pm$6  &  11$\pm$3 \\
3 &  36$\pm$9   &  \nodata     &  46$\pm$12  & 
	\nodata     &  71$\pm$18  &  12$\pm$3  &  7.0$\pm$1.8 \\
4 &  27$\pm$7   &  \nodata     &  56$\pm$14  & 
	\nodata     &  74$\pm$19  &  30$\pm$7  &  13$\pm$3 \\
\enddata
\tablecomments{Line peak temperatures labeled as T$^X_Y$, where $X$ is
the carbon isotope and Y is the rotational transition. All values are
in Kelvins.}
\tablenotetext{a}{See labeled positions in Figure~\ref{map}.}
\tablenotetext{b}{Data from this work.}
\tablenotetext{c}{Data from \citetalias{wil01}.}
\tablenotetext{d}{Data from K.~N.~Allers 2005, in preparation.}
\tablenotetext{e}{Data from \citet{tigg93}.}
\tablenotetext{f}{Data from \citet{plu00}.}
\end{deluxetable}

\begin{deluxetable}{ccccc}
\tablecolumns{5}
\tablewidth{0pt}
\tablecaption{Gas parameters derived from LVG model\label{fits}}
\tablehead{\colhead{Position} & \colhead{T$_{kin}$} & 
\colhead{log $n$} & \colhead{log($dN/dv$)} & $\chi^2_{r,min}$ \\
\colhead{} & \colhead{(K)} & 
\colhead{(\cc)} & \colhead{(\dndv)}}
\startdata
1 & 120$\pm$20 & $3.5-4.6$\tablenotemark{a} & $18.0\pm0.2$ & 0.83 \\
2 & 120$\pm$20 & $3.5-4.3$\tablenotemark{a} & $18.0\pm0.2$ & 0.81 \\
3 &  70$\pm$10 & $4.1\pm0.3$ & $17.7\pm0.2$ & 1.46 \\
4 &  70$\pm$10 & $3.8\pm0.3$ & $18.1\pm0.2$ & 1.16 \\
\enddata
\tablenotetext{a}{Poorly constrained upper limit. See text for details.}
\end{deluxetable}

\end{document}